\pdfoutput=1

\documentclass[letterpaper]{article}
\usepackage{iccc}
\usepackage{xcolor}
\usepackage{url}
\usepackage{amsmath}

\usepackage{epsfig}
\usepackage{graphicx}
\usepackage{wrapfig}
\usepackage[belowskip=1pt,aboveskip=5pt,font=small]{caption}
\usepackage[belowskip=0pt,aboveskip=3pt,font=small]{subcaption}
\usepackage[T1]{fontenc}
\usepackage{times}
\usepackage[scaled=0.90]{inconsolata}

\def\code#1{\texttt{#1}}

\title{Lemotif: An Affective Visual Journal Using Deep Neural Networks}
\author{
    X. Alice Li\\
    Georgia Tech \\
    \texttt{xali@gatech.edu}
    \And
    Devi Parikh\\
    Facebook AI Research \& Georgia Tech \\
    \texttt{parikh@gatech.edu}
}
\begin{document} 
\maketitle

\begin{abstract}
\begin{quote}

We present Lemotif, an integrated natural language processing and image generation system that uses machine learning to (1) parse a text-based input journal entry describing the user's day for salient themes and emotions and (2) visualize the detected themes and emotions in creative and appealing image motifs. Synthesizing approaches from artificial intelligence and psychology, Lemotif acts as an affective visual journal, encouraging users to regularly write and reflect on their daily experiences through visual reinforcement. By making patterns in emotions and their sources more apparent, Lemotif aims to help users better understand their emotional lives, identify opportunities for action, and track the effectiveness of behavioral changes over time. We verify via human studies that prospective users prefer motifs generated by Lemotif over corresponding baselines, find the motifs representative of their journal entries, and think they would be more likely to journal regularly using a Lemotif-based app. 

\end{quote}
\end{abstract}

\section{Introduction}

Our emotional well being is important. In part due to its subjective nature, it is difficult to find patterns in what we feel, how often we feel it, and what the source of those feelings tends to be. Without this assessment, it is difficult to tweak our choices to optimize our emotional well being.

Meanwhile, innovations in artificial intelligence have produced powerful neural networks capable of sophisticated analytic and generative tasks.  There exists great potential for machine learning to address human-centered needs, using creative interdisciplinary approaches to model subjective qualities like emotion which can be difficult to quantify. 

In this paper we introduce Lemotif, an integrated natural language processing (NLP) and image generation system serving as an affective visual journal. Given a text-based journal entry describing aspects of the user's day, a multi-label classifier building upon the Bidirectional Encoder Representations from Transformers (BERT) language model \cite{devlin2018bert} extracts salient topics and associated emotions from the provided input. An image generation algorithm then creates motifs conditioned upon the detected topics and emotions; we offer several representation styles, including a neural network trained on abstract art.

\begin{figure}
	\centering
	\begin{subfigure}{1\columnwidth}
	\footnotesize{
	\textbf{Sample journal entries collected:}
	
	Had a wonderful and insightful conversation with my best friend. Many of the things we talked about we were on the same wavelength ... Made a new recipe today and it turned out tasting really well. Will have to remember to put it into my repertoire more often ... Had a terrible time getting comfortable when trying to sleep last night. Was so tired today I tried a nap and that didn't go well either.
\medbreak{}
\textbf{Sample motifs generated:}
}
		\caption*{}
		\vspace{-10pt}
	\end{subfigure}
	\begin{subfigure}{1\columnwidth}
		\centering
        \includegraphics[width=1\columnwidth]{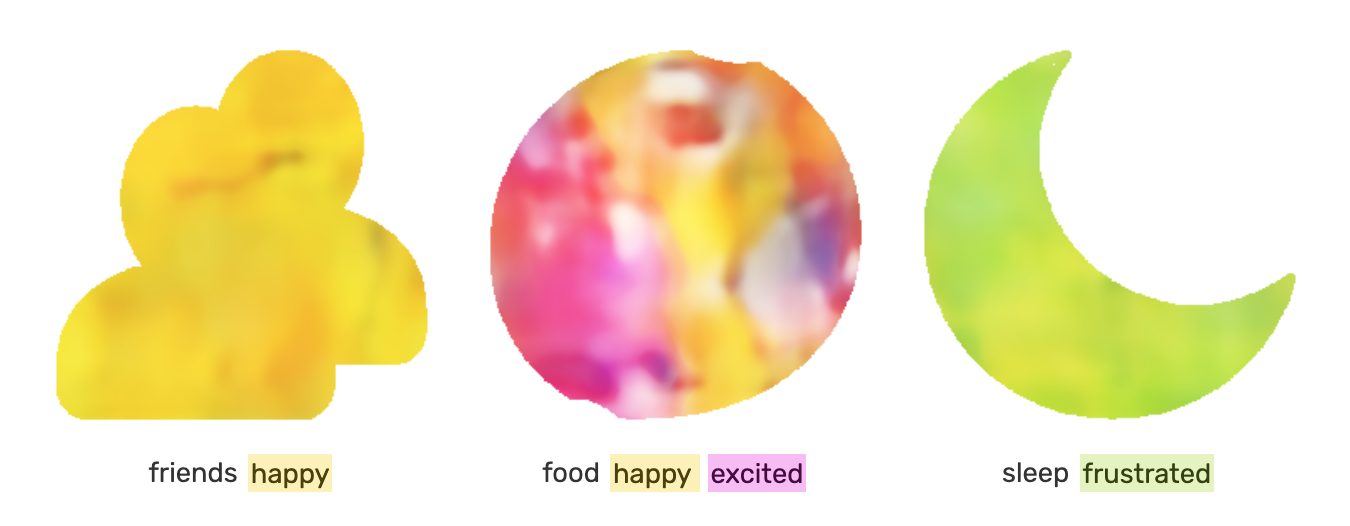}
	\end{subfigure}
	\caption{Lemotif uses machine learning to extract and visualize salient themes and emotions from text. We hypothesize that the creative aspects of Lemotif have the potential to make journaling more actionable and engaging.} 
    \label{fig:teaser}
    \vspace{-15pt}
\end{figure}

The concrete instantiation of Lemotif is shown in Fig.~\ref{fig:teaser}. The core principles behind our approach are:
(1) The generated motif separately depicts each salient topic to make the source of feelings visually apparent to the user. 
(2) The generated motif depicts topics visually using outline shapes as seen in Fig.~\ref{fig:teaser} so the feeling-topic association is more apparent and better grounded in the user's mind.
(3) The generated motif depicts emotions visually as well, using color mappings as seen in Fig.~\ref{fig:teaser}.
(4) The generated motif is creative and attractive to provide visual reinforcement for user engagement. (5) The overall system uses machine intelligence to automate text analysis and image synthesis, allowing users to write naturally and receive computationally generated analysis of typical journal entry inputs.

We evaluate Lemotif qualitatively via human studies assessing (1) whether the topic-shape and feeling-color mappings are meaningful to users, (2) whether subjects favor the generated motifs over corresponding baselines, (3) whether subjects consider the generated motifs representative of their journal entries, and (4) whether subjects would engage positively with such an app, including being willing to use an app like Lemotif and feeling like the app would encourage them to journal more regularly.  The NLP model is evaluated as a multi-label classifier calculating F1 and normalized accuracy through cross-validation. We report favorable results on all fronts. Our code and trained models are available at \texttt{https://github.com/xaliceli/lemotif}. A demo is available at \texttt{http://lemotif.cloudcv.org}.

\section{Related Work}
\label{sec:rw}
Our work processes text to extract key topic and emotion labels, maps these abstract concepts to visual entities (shape and color), and generates a visual depiction of the input text according to the extracted labels. In this section we discuss prior work relating to each individual component as well as our overall goal of creatively summarizing journal entries.

\subsubsection{Journaling Tools} Our work is motivated by psychological research indicating that writing about emotions can support mental health ~\cite{pennebaker1997writing}. Most existing journaling tools and apps allow users to log their lives without focusing on identifying themes or patterns. Emphasis is often on easy incorporation of external content, including multimedia, hand annotations, maps, and search tools. Our focus is more on making associations between a user's feelings and aspects of their life apparent. When journaling apps claim to be `visual' (e.g., HeyDay), they typically refer to allowing visual input modalities such as images and videos. Our work produces a visual modality as an output. Life Calendar (\texttt{journallife.me}) comes closest to our approach, showing a single-colored dot (red, yellow, or green) for each week that captures the mood of the user in that week (negative, neutral, positive). This allows one to find correlations between time of month or year and emotion (e.g., happier in the summer). But it does not help identify sources of nuanced emotions on a day-to-day basis. In our experiments, we compare our motifs to a visualization that mimics this and find that subjects strongly prefer our nuanced and creative visualizations. 

\subsubsection{Natural Language Processing} Our task of identifying topics and emotions from text is related to existing work on keyword extraction and sentiment analysis, though sentiment analysis is commonly approached as a binary ("positive" vs "negative") or trinary ("positive", "neutral", "negative") problem. Recently, BERT ~\cite{devlin2018bert} and GPT-2 ~\cite{radford2019language} have successfully pre-trained NLP models on large unlabeled text datasets, learning language representations that can then be generalized to downstream tasks. We fine-tune BERT (pre-trained on the BooksCorpus and English Wikipedia datasets) on a custom dataset for our specific task of identifying up to 11 topics and up to 18 associated emotions, a form of aspect-based sentiment analysis ~\cite{pontiki2016semeval}.

\subsubsection{Visual Representation} Our work draws upon existing research on common associations between colors and emotions ~\cite{emo_to_color}. Studies have also indicated associations between the visual qualities of lines and the emotions they evoke ~\cite{poffenberger1924feeling}. There exists fascinating work on projecting a spectrum of human emotions on an interactive map with associated video clips that elicit those emotions~\cite{emo_mapping_video} and to audio gasps that people make when expressing those emotions~\cite{emo_mapping_audio}. ~\cite{emotion_from_paintings} studied the emotions evoked in viewers of paintings generated by a creative AI system, while ~\cite{AlvarezMelis2017TheEG} collected emotional labels for artworks and trained a generative adversarial network to synthesize new images conditioned upon emotional labels. We extend the foundational idea that emotions can be represented in visual form by using a relatively rich set of 18 nuanced emotions as a core design principle of Lemotif. The use of recognizable icons to represent topics is also a common feature in popular note-taking apps such as Notion (\texttt{notion.so}); we take a similar approach, using shapes to represent topics in Lemotif.

\subsubsection{Image Synthesis} Approaches in multi-modal AI that generate natural images from their language descriptions~\cite{scott_reed_text2im}  using generative adversarial networks (GANs)  ~\cite{Goodfellow2014GenerativeAN} are relevant. Convolutional neural networks comprising an encoder block for feature extraction and a decoder block for synthesis and reconstruction, or an "autoencoder" model ~\cite{hinton2006reducing}, have been widely used for image-to-image translation tasks, including image super-resolution taking small samples as inputs and reconstructing larger outputs with greater detail ~\cite{dong2015image}. We take this super-resolution approach for our autoencoder visualization style, with more details in the following section. We also draw from the tradition of generative art ~\cite{galanter2003generative} in our geometric visualization styles, using computational methods and mathematical principles for aesthetic purposes.

\section{Approach}
\label{sec:approach}

\begin{figure*}
	\centering
	\begin{subfigure}{0.18\columnwidth}
		\centering
		\includegraphics[width=1\columnwidth]{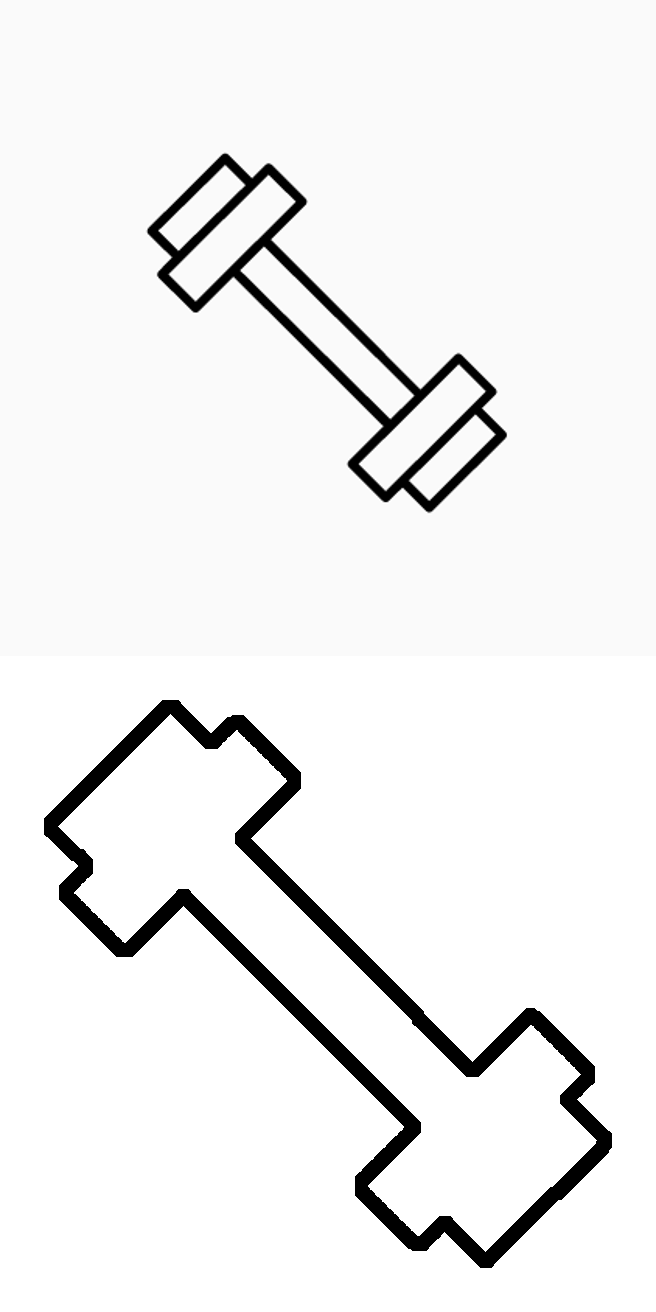}
		\caption*{Exercise}
	\end{subfigure}
	\begin{subfigure}{0.18\columnwidth}
		\centering
        \includegraphics[width=1\columnwidth]{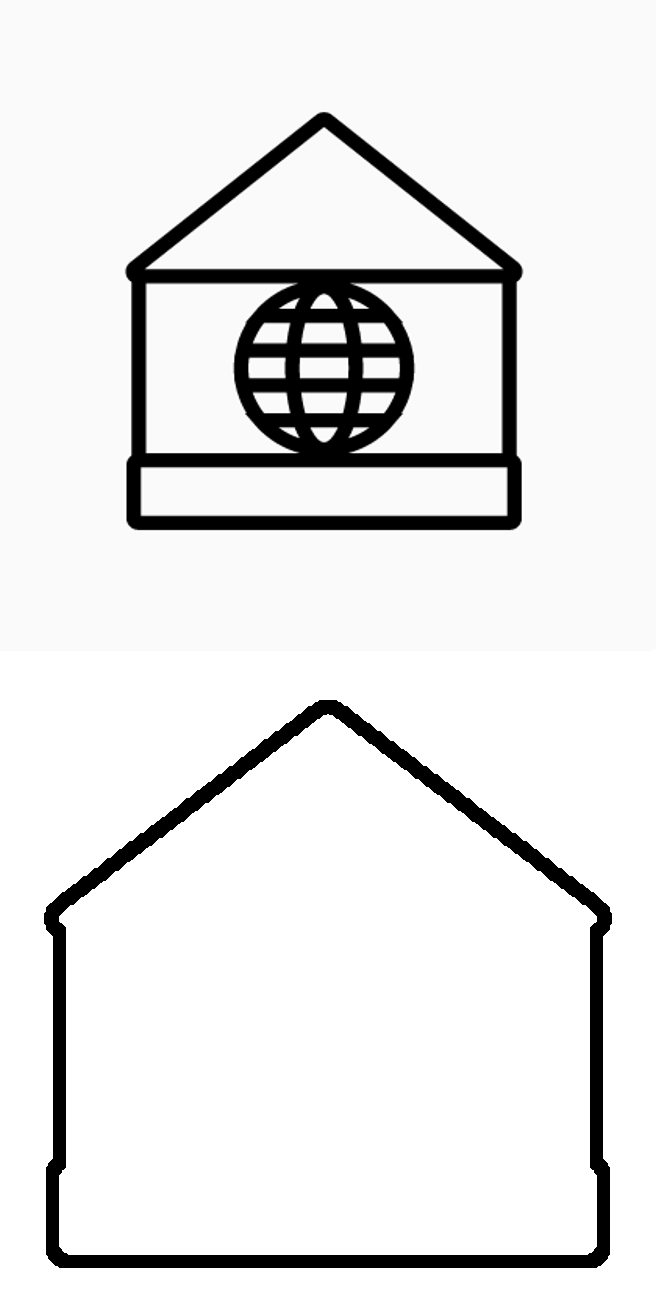}
		\caption*{Family}
	\end{subfigure}
	\begin{subfigure}{0.18\columnwidth}
		\centering
		\includegraphics[width=1\columnwidth]{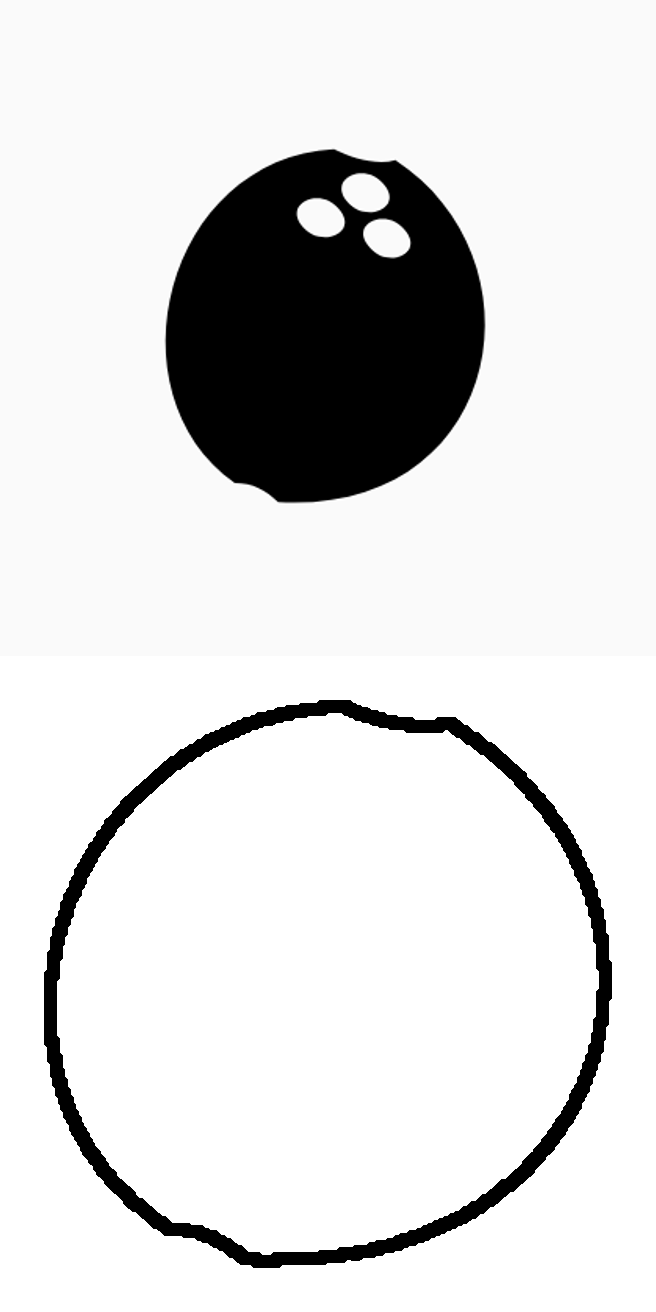}
		\caption*{Food}
	\end{subfigure}
	\begin{subfigure}{0.18\columnwidth}
	    \centering
		\includegraphics[width=1\columnwidth]{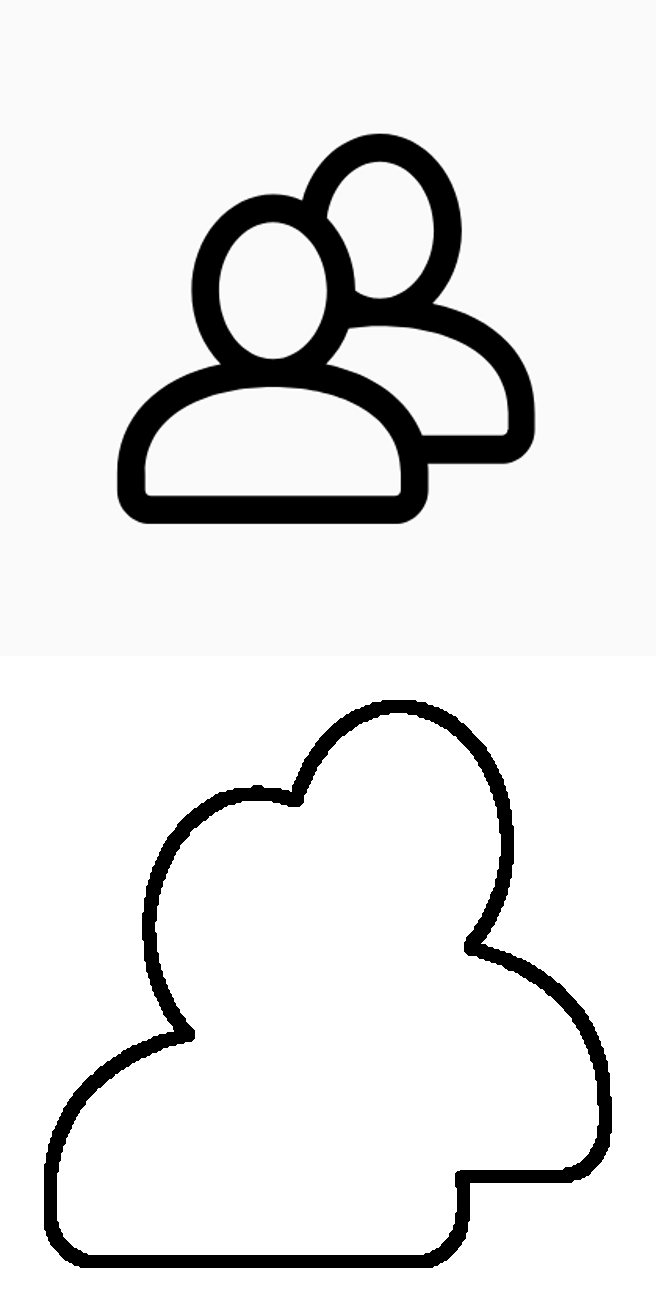}
		\caption*{Friends}
	\end{subfigure}
	\begin{subfigure}{0.18\columnwidth}
		\centering
		\includegraphics[width=1\columnwidth]{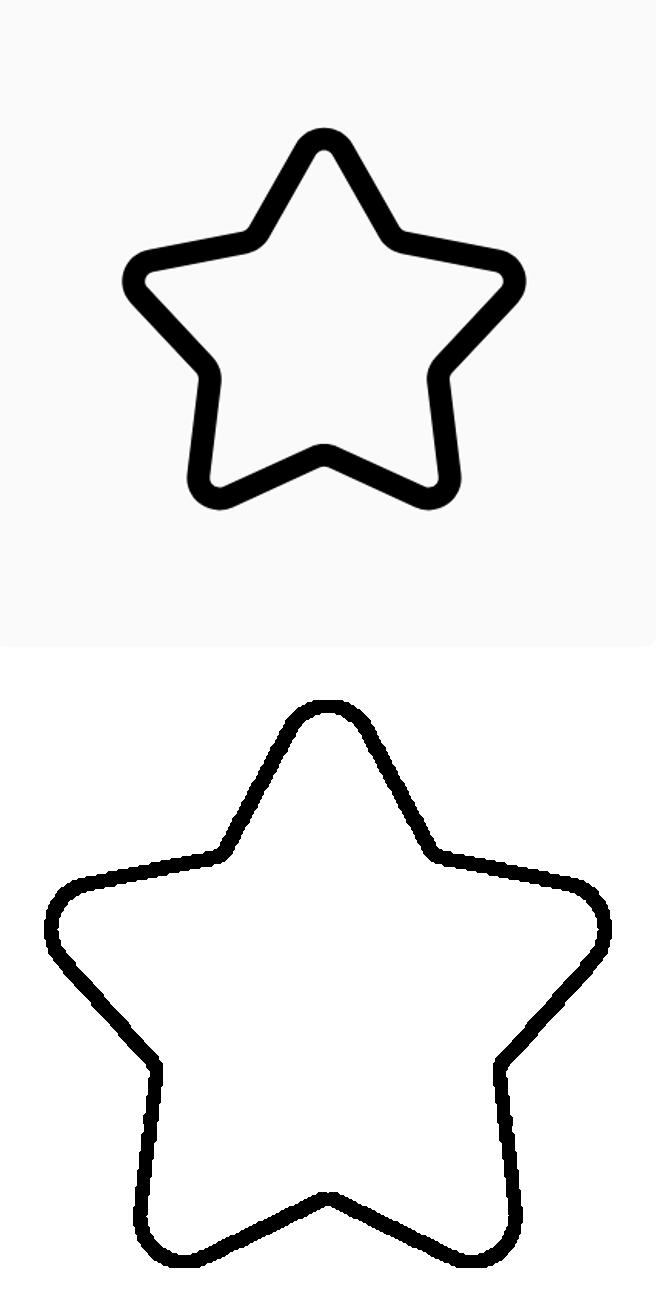}
		\caption*{God}
	\end{subfigure}
	\begin{subfigure}{0.18\columnwidth}
		\centering
        \includegraphics[width=1\columnwidth]{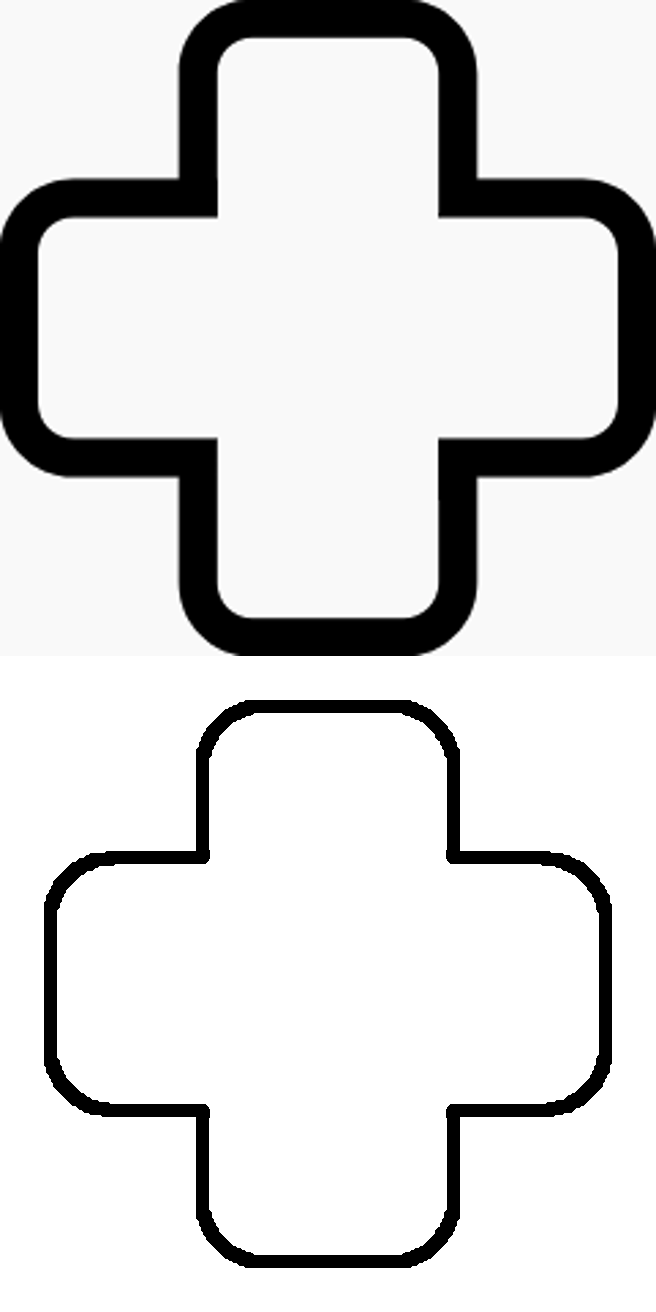}
		\caption*{Health}
	\end{subfigure}
		\begin{subfigure}{0.18\columnwidth}
	    \centering
		\includegraphics[width=1\columnwidth]{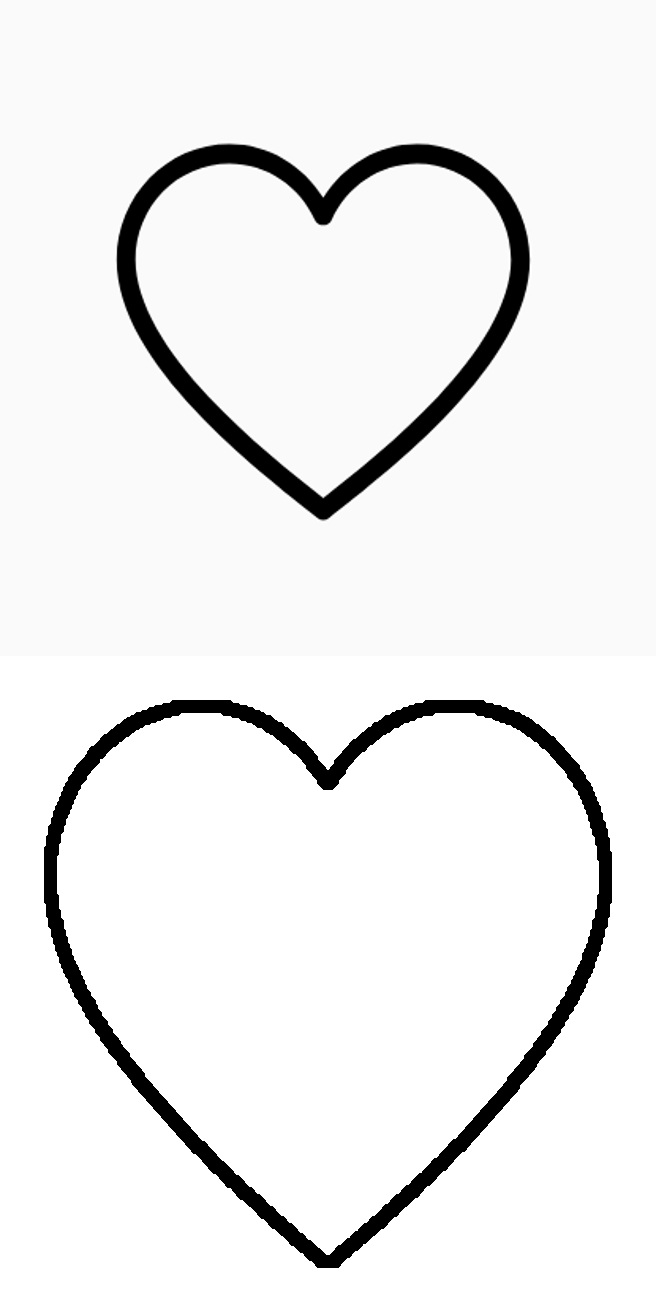}
		\caption*{Love}
	\end{subfigure}
	\begin{subfigure}{0.18\columnwidth}
		\centering
		\includegraphics[width=1\columnwidth]{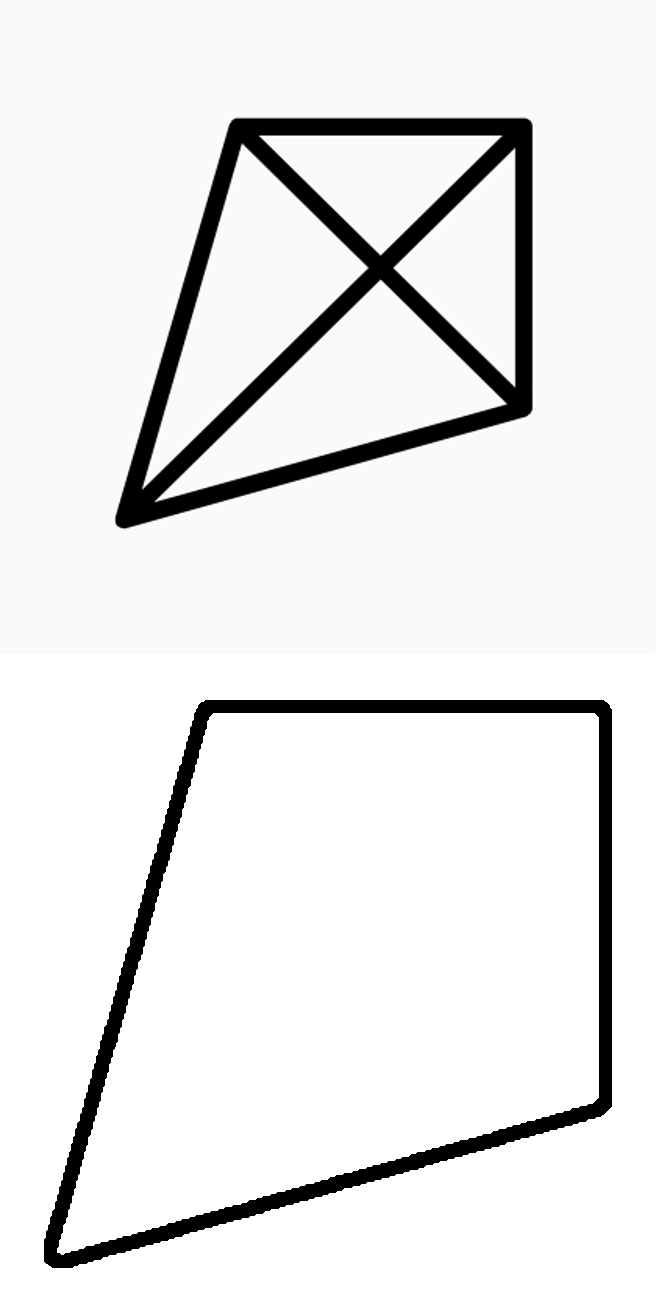}
		\caption*{Recreation}
	\end{subfigure}
	\begin{subfigure}{0.18\columnwidth}
		\centering
        \includegraphics[width=1\columnwidth]{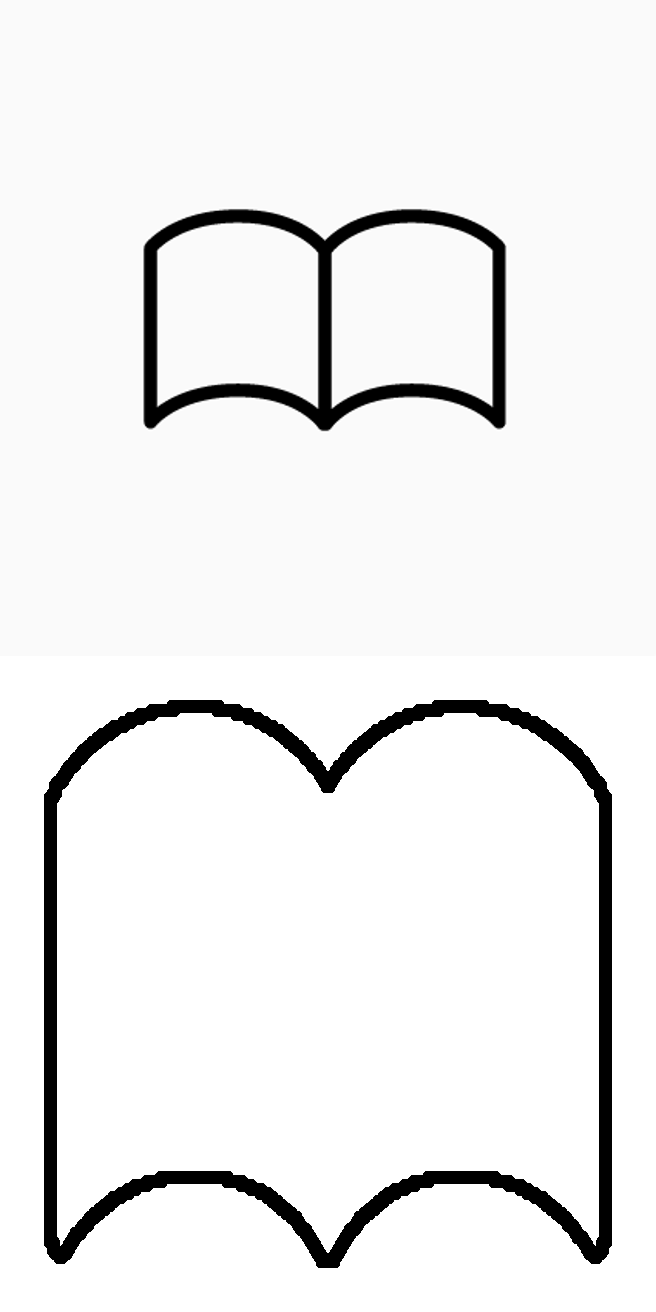}
		\caption*{School}
	\end{subfigure}
		\begin{subfigure}{0.18\columnwidth}
		\centering
        \includegraphics[width=1\columnwidth]{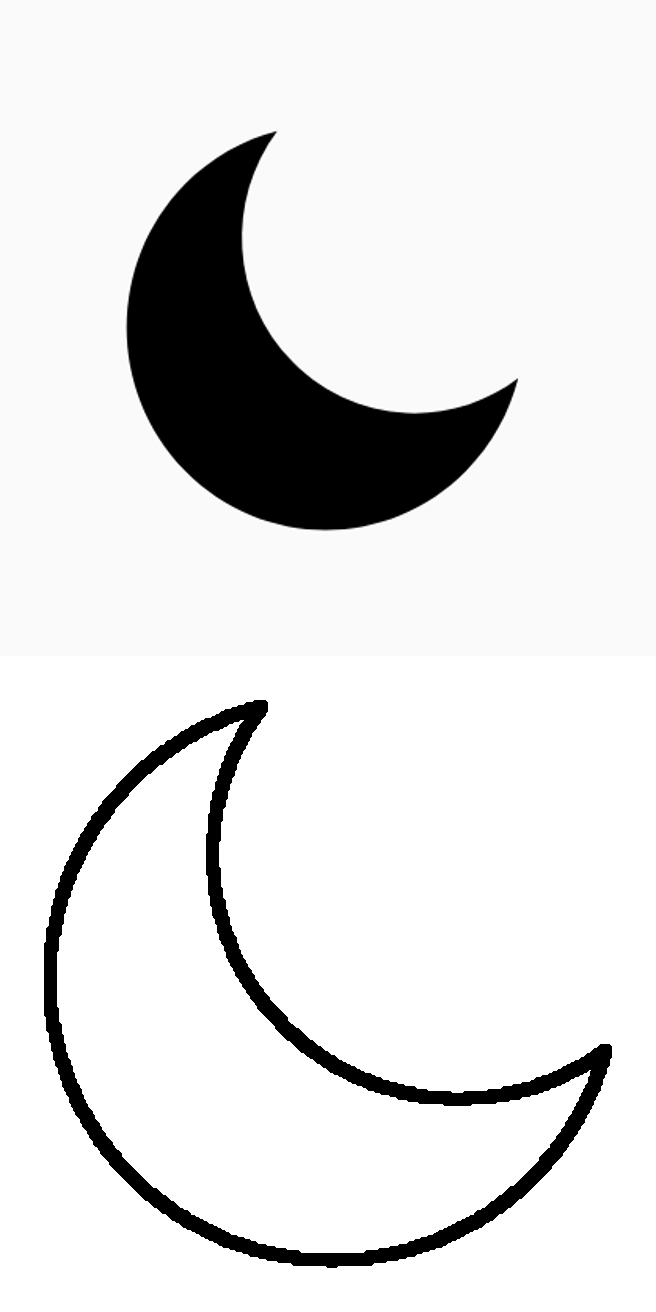}
		\caption*{Sleep}
	\end{subfigure}
		\begin{subfigure}{0.18\columnwidth}
		\centering
        \includegraphics[width=1\columnwidth]{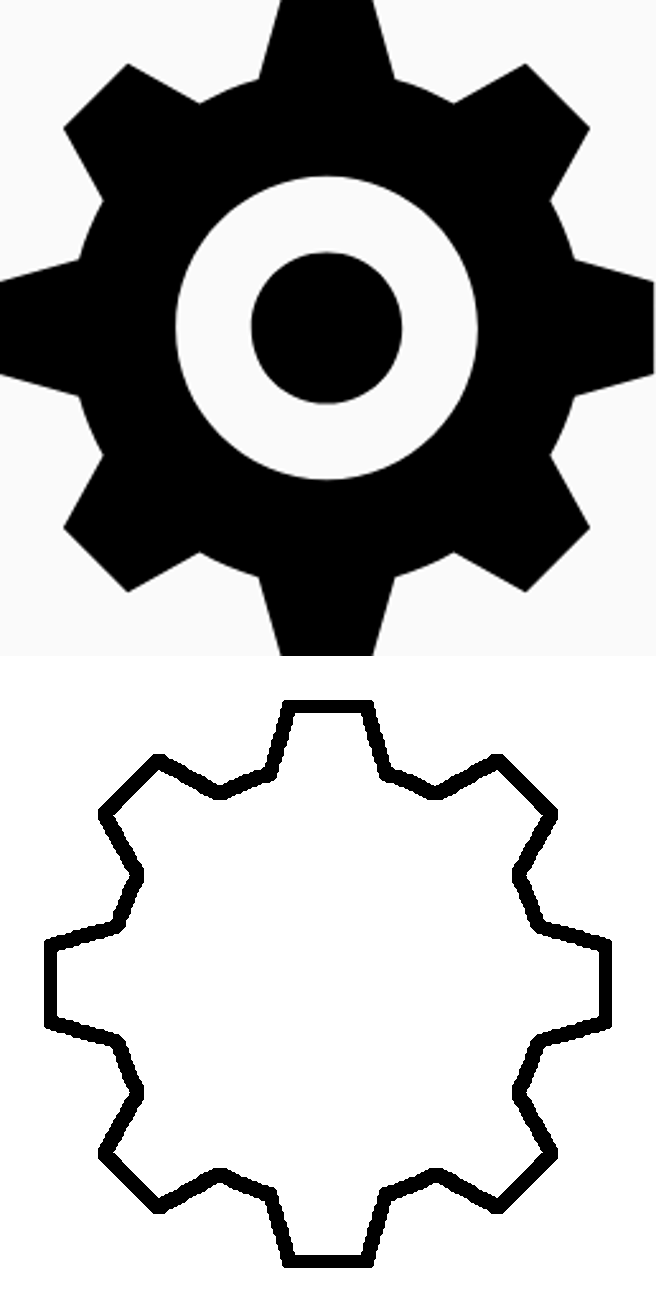}
		\caption*{Work}
	\end{subfigure}
	\caption{Icons used to represent various topics. We select icons from the Noun Project (top) and process them to binarize, resize and recenter before extracting the outer shape (bottom).}
    \label{fig:icons}
        \vspace{-10pt}
\end{figure*}
\begin{figure}[t!]
    \centering
    \includegraphics[width=\columnwidth]{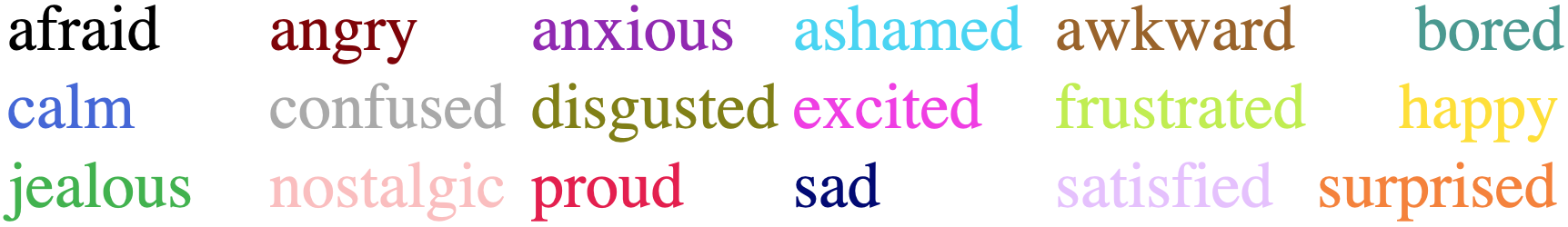}
    \caption{Colors used to represent various feelings or emotions.}
    \label{fig:colors}
    \vspace{-15pt}
\end{figure}

Below we describe our approach to processing journal entries, mapping concepts to visual content, and generating representative motifs.

\subsection{Natural Language Processing}

Our NLP objective is to take free-form text input and predict salient topic and emotion labels. To that end, we fine-tune BERT to serve as a multi-label classifier. We use the BERT-Base model containing 12 encoder layers, 768 hidden units per layer, and 12 attention heads for a total of 110M parameters \cite{devlin2018bert}. To BERT-Base we append a fully-connected multi-layer perceptron containing 768 hidden units with sigmoid activation and 29 output nodes corresponding to our 11 topics and 18 emotions. We fine-tune this model on our dataset of text samples with user-annotated labels (more details in the Dataset section), optimizing over sigmoid cross-entropy loss. Labels above a set probability threshold are returned as the salient topics and associated emotions; we use 0.2 as our threshold, chosen through cross-validation. These labels are then used as inputs for the image generation algorithms, such that each motif represents one topic with the highest probability and a set of up to four emotions with the highest probabilities.

\subsection{Topics and Emotions}

Human experiences are complex, multimodal, and  subjective. A system that identifies and visualizes abstract content about an individual's emotional life must be both comprehensive and intelligible, addressing most common themes in life through discrete labels while representing this information in a format humans recognize and approve of. Below we outline our approach to identifying our target labels and mapping these concepts to visual representations.

\subsubsection{Topics}

The 11 topics in our pre-defined list are shown in Fig.~\ref{fig:icons}. This list was determined by a mix of brainstorming and searching online for what topics users typically talk about in their journals. As part of our evaluation, we asked survey participants in an Amazon Mechanical Turk (AMT) study if they felt a topic they would like to talk about was missing. 99 subjects out of 100 said this list was sufficient. One user suggested adding pets as a topic.

\subsubsection{Emotions}

The 18 emotions in our pre-defined list are shown in Fig.~\ref{fig:colors}. This list was curated from~\cite{emo_mapping_video} and our assessment of what emotions are likely on a day-to-day basis. Again, as part of our evaluation, we asked users from the same AMT study described above if they felt an emotion they would like to talk about was missing. All 100 subjects said the list was sufficient.

\subsubsection{Shapes for topics}

Lemotif uses a pre-defined mapping from topics to visual icons of shapes depicting that topic. These are shown in Fig.~\ref{fig:icons}. To identify our list of icons, we started with The Noun Project (\code{http://thenounproject.com}) which contains over two million binary icons created by designers all over the world. We searched The Noun Project for each of the topics to ensure that the icons we pick are relevant to the topic (e.g., book for school). From the relevant icons, we selected those that are not visually complex so the generated motif is clear. We automatically binarize the image, crop the icon, and resize it to a canonical size. To further simplify the icons, we post-process them to retain only their outer shape and discard the inner details. This was done by keeping only the extreme points of the shape in each row and column of the image, providing a thin and sparse outline of the icon. For completeness, we dilate the sparse outline using morphological filtering. The resulting icons are shown in the bottom row of Fig.~\ref{fig:icons}. 

\subsubsection{Colors for emotions}

Lemotif uses a pre-defined mapping from emotions to corresponding colors associated with that emotion, as shown in Fig.~\ref{fig:colors}. These colors were selected based on common associations (e.g., dark red for angry) as indicated by studies~\cite{emo_to_color} while making sure that each color is visually distinct ~\cite{distinctness}.

\subsection{Image Synthesis}

Taking a set of labels extracted by the NLP model consisting of topics and emotions, Lemotif generates image motifs depicting these salient themes in visual form. Acknowledging that creative preferences are inherently subjective and individual, we offer six creative visualization styles described next. The generated visualization image is then bounded by a shape icon representing the relevant topic. The human user exercises creative input in selecting a motif style and adjusting various input parameters according to personal taste, while the algorithm produces unique motifs with stochastic variations for each generated image. 

\subsubsection{Autoencoder} 
\begin{figure*}[ht]
	\centering
	\begin{subfigure}{0.66\columnwidth}
		\centering
		\includegraphics[width=1\columnwidth]{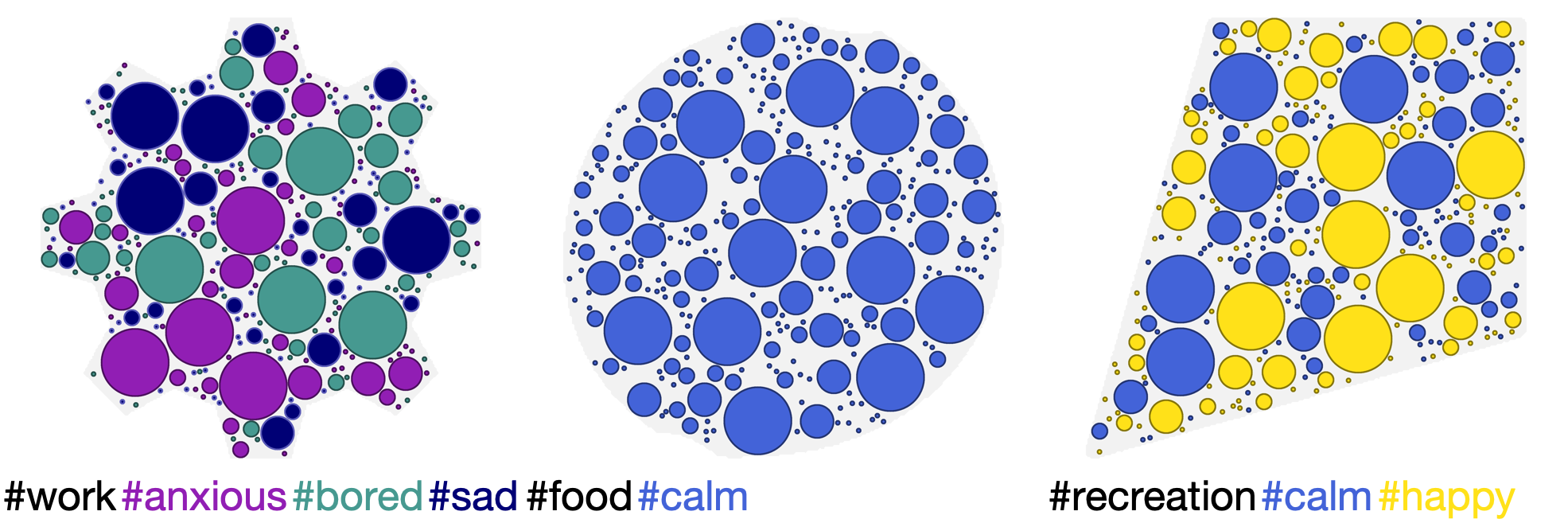}
		\caption*{A1 - Circle Packing}
	\end{subfigure}
	\begin{subfigure}{0.66\columnwidth}
		\centering
        \includegraphics[width=1\columnwidth]{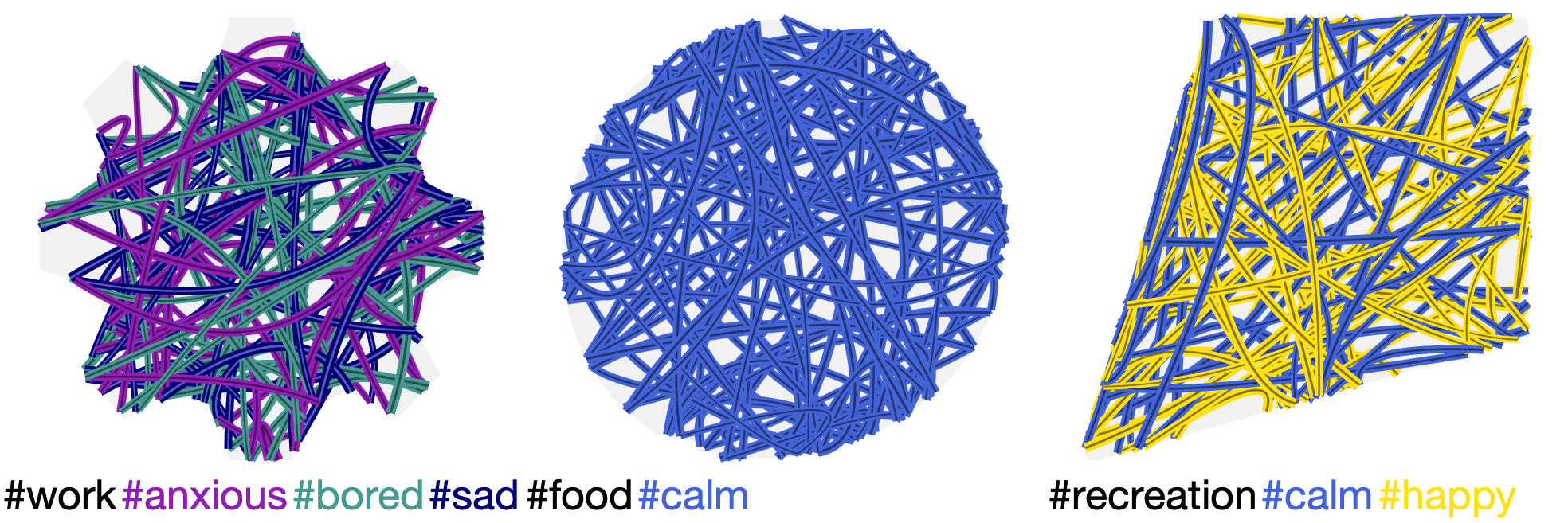}
		\caption*{A2 - String Doll}
	\end{subfigure}
		\begin{subfigure}{0.66\columnwidth}
		\centering
        \includegraphics[width=1\columnwidth]{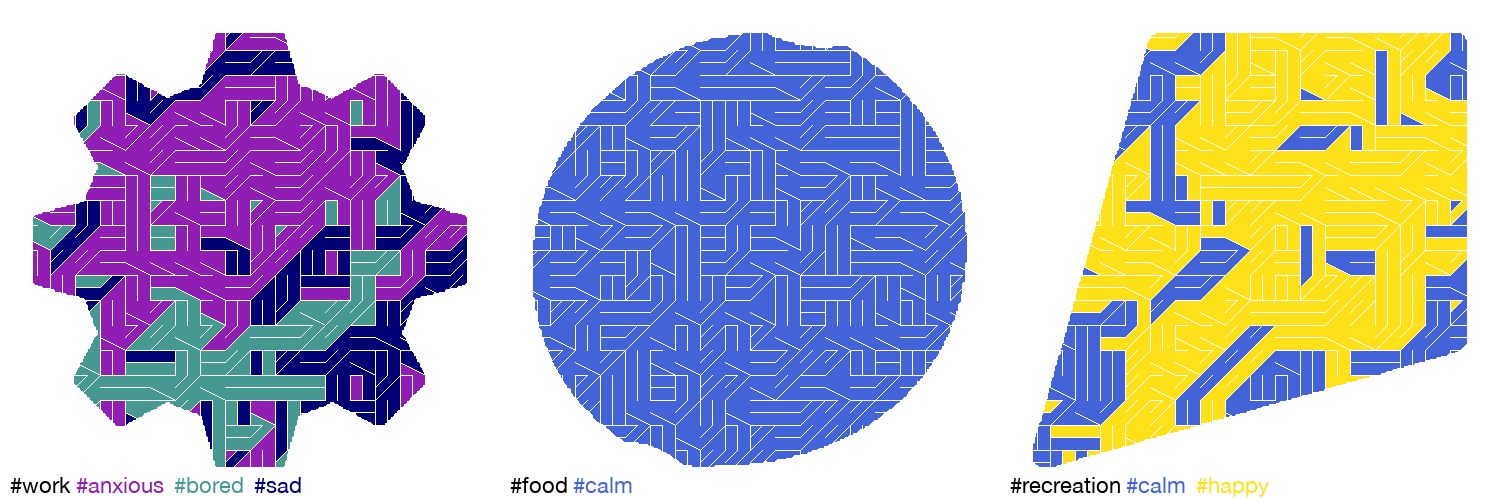}
		\caption*{A3 - Carpet}
	\end{subfigure}
		\begin{subfigure}{0.66\columnwidth}
		\centering
        \includegraphics[width=1\columnwidth]{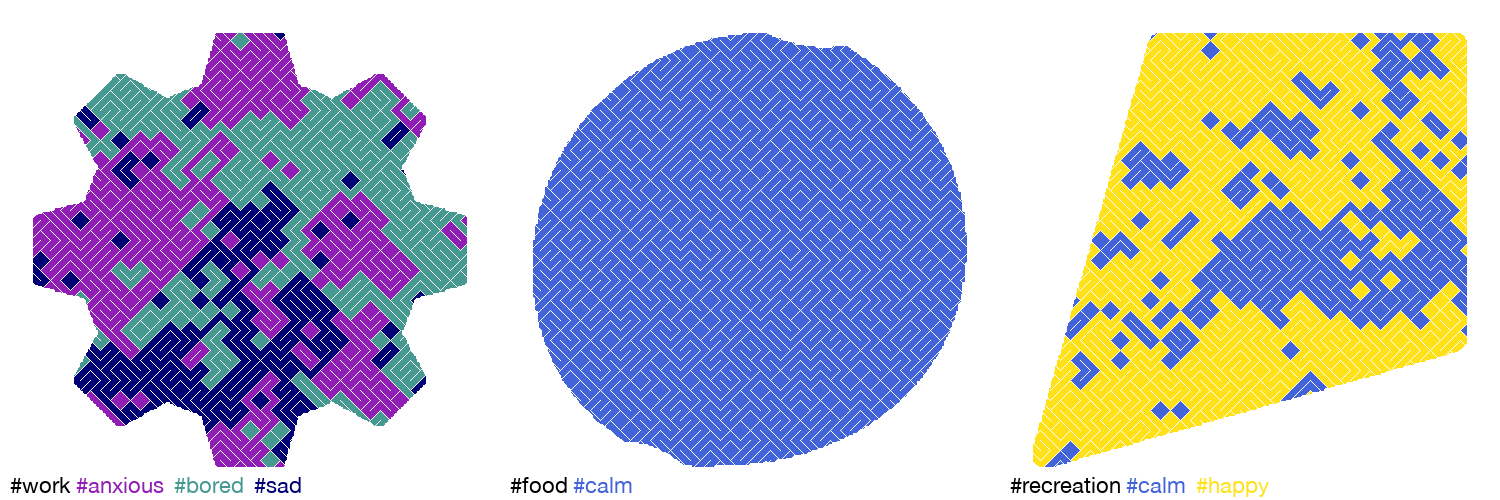}
		\caption*{A4 - Tile}
	\end{subfigure}
		\begin{subfigure}{0.66\columnwidth}
		\centering
        \includegraphics[width=1\columnwidth]{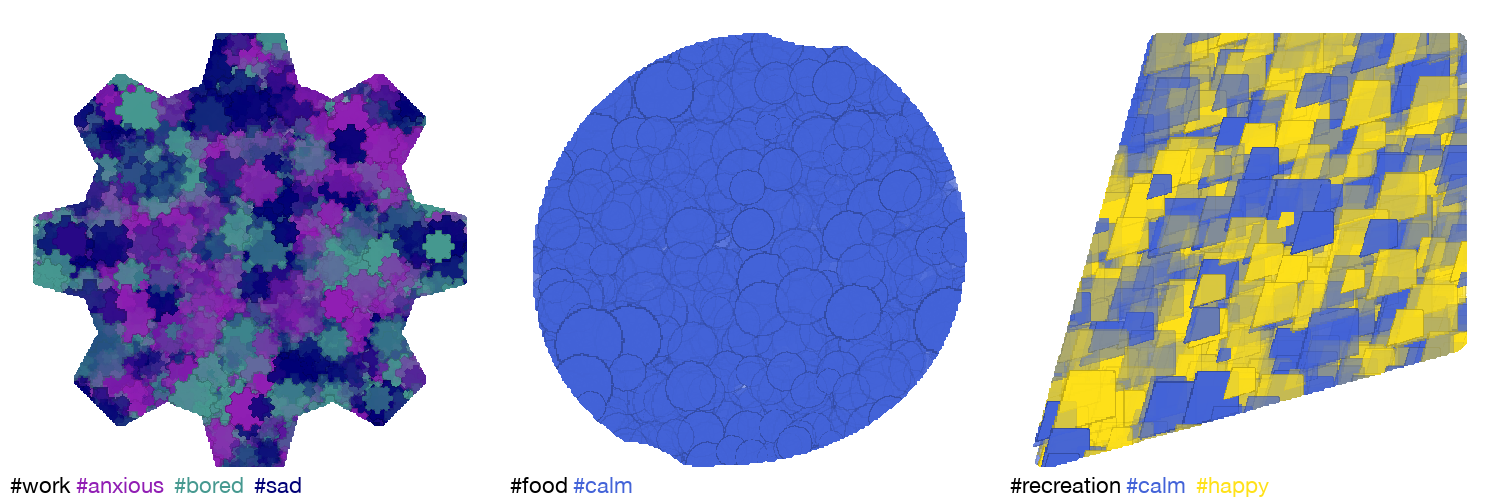}
		\caption*{A5 - Glass}
	\end{subfigure}
		\begin{subfigure}{0.66\columnwidth}
		\centering
        \includegraphics[width=1\columnwidth]{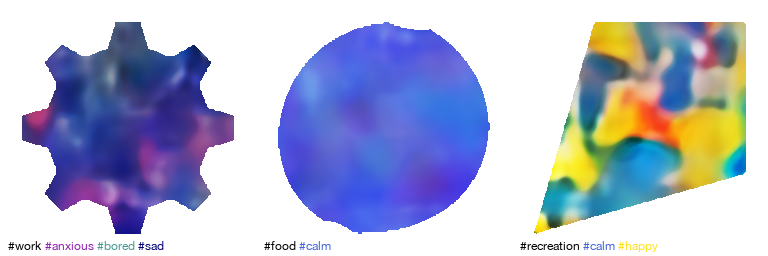}
		\caption*{A6 - Autoencoder}
	\end{subfigure}
	\begin{subfigure}{0.66\columnwidth}
		\centering
		\includegraphics[width=1\columnwidth]{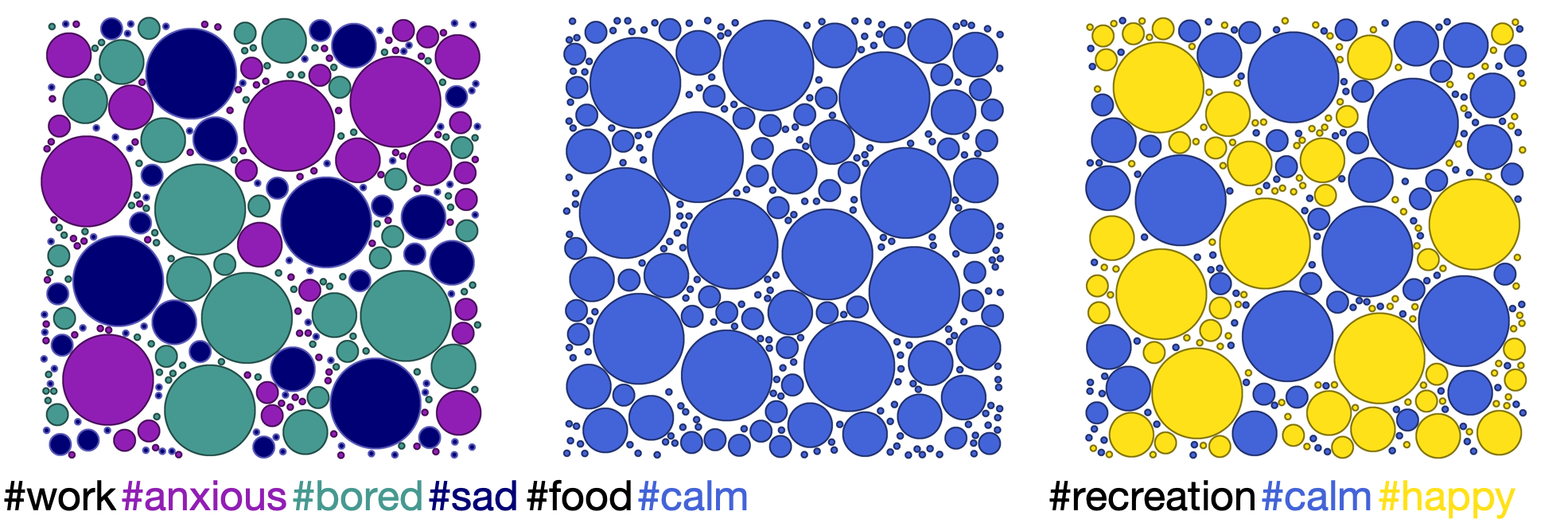}
		\caption*{B1}
	\end{subfigure}
	\begin{subfigure}{0.66\columnwidth}
	    \centering
		\includegraphics[width=1\columnwidth]{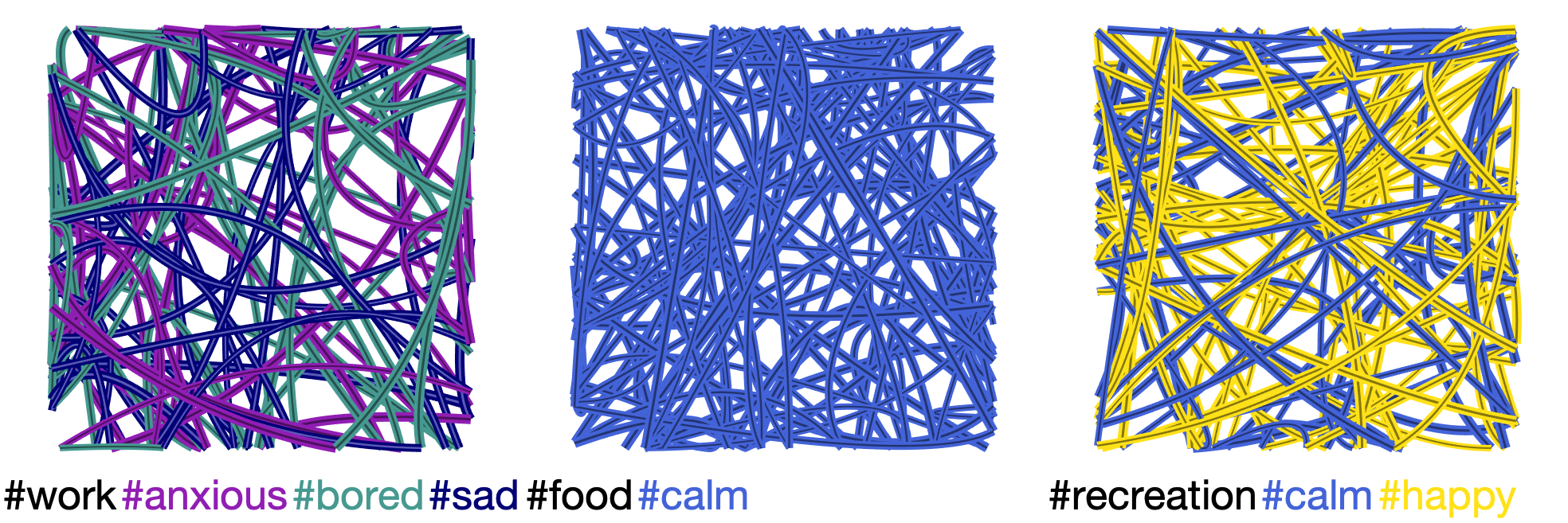}
		\caption*{B2}
	\end{subfigure}
	\begin{subfigure}{0.66\columnwidth}
		\centering
		\includegraphics[width=1\columnwidth]{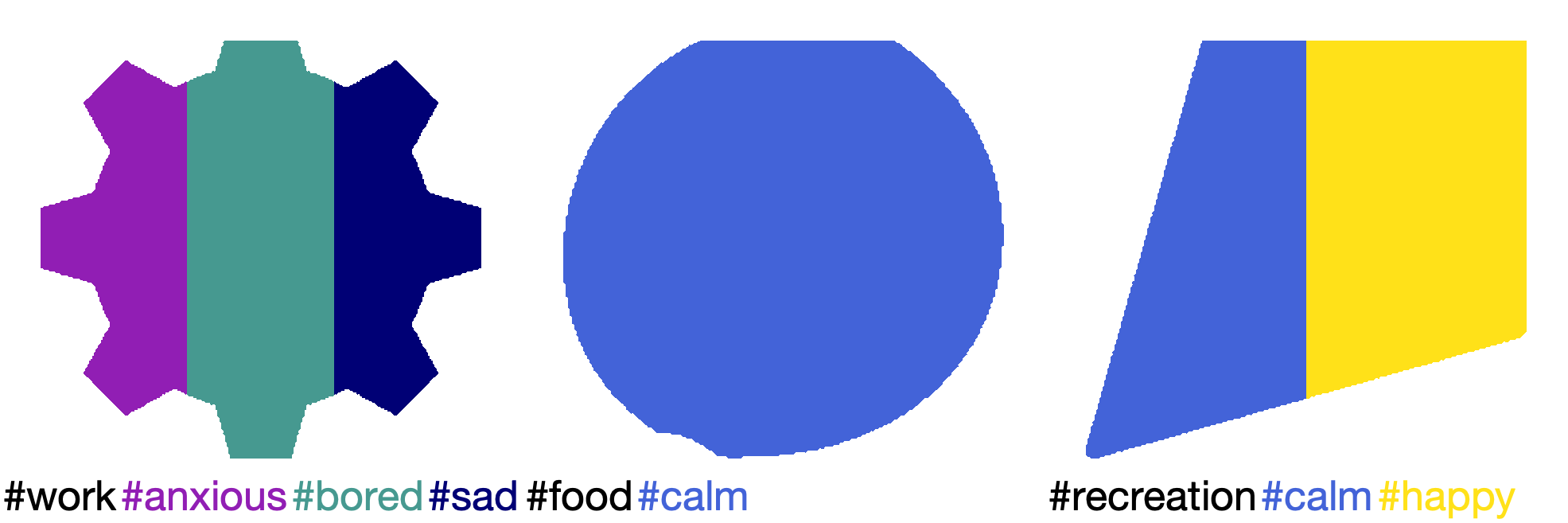}
		\caption*{B3}
	\end{subfigure}
	\begin{subfigure}{0.58\columnwidth}
		\centering
        \includegraphics[width=1\columnwidth]{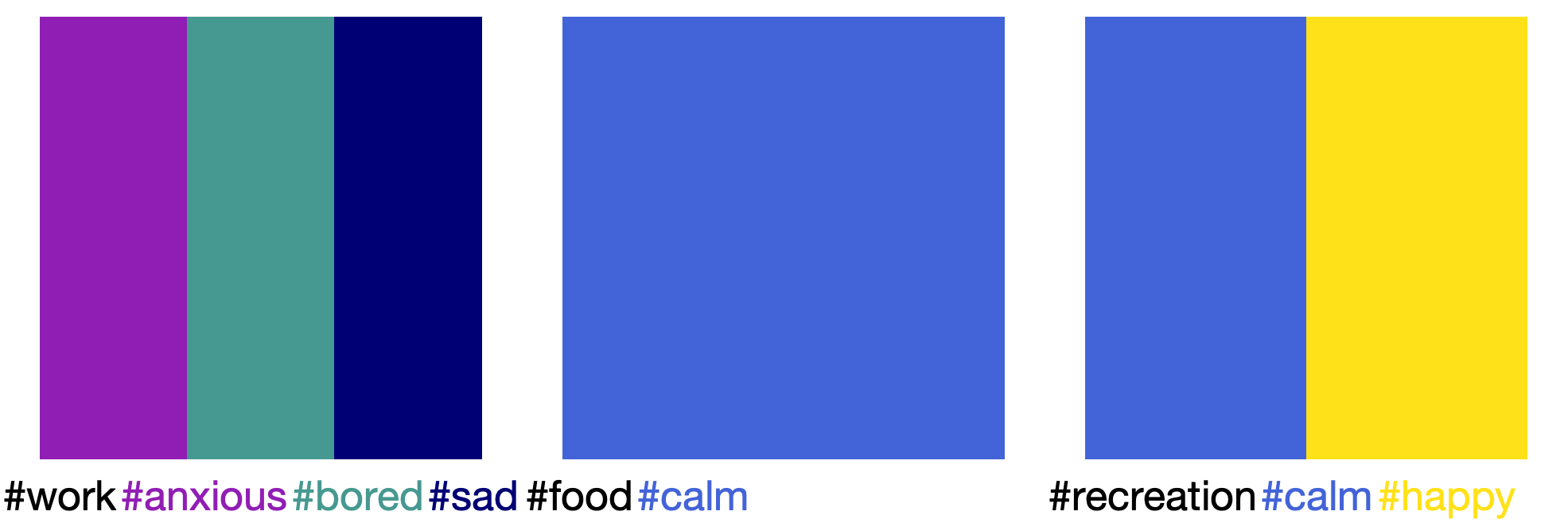}
		\caption*{B4}
	\end{subfigure}
		\begin{subfigure}{0.58\columnwidth}
	    \centering
		\includegraphics[width=1\columnwidth]{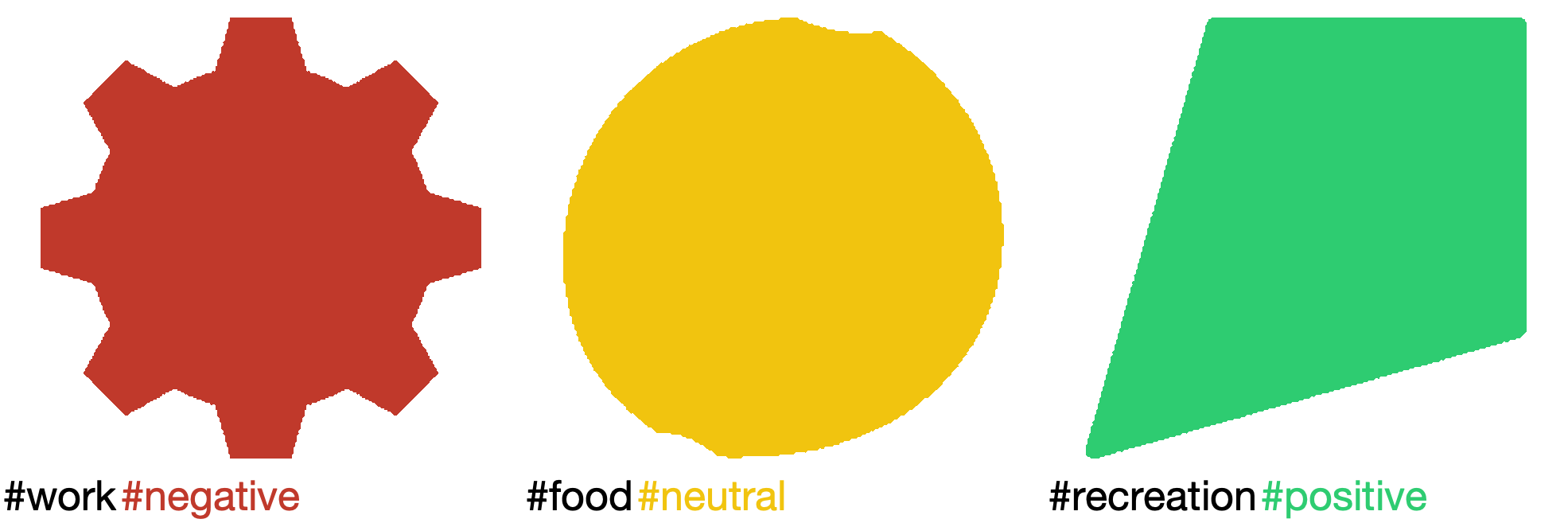}
		\caption*{B5}
	\end{subfigure}
	\begin{subfigure}{0.58\columnwidth}
		\centering
		\includegraphics[width=1\columnwidth]{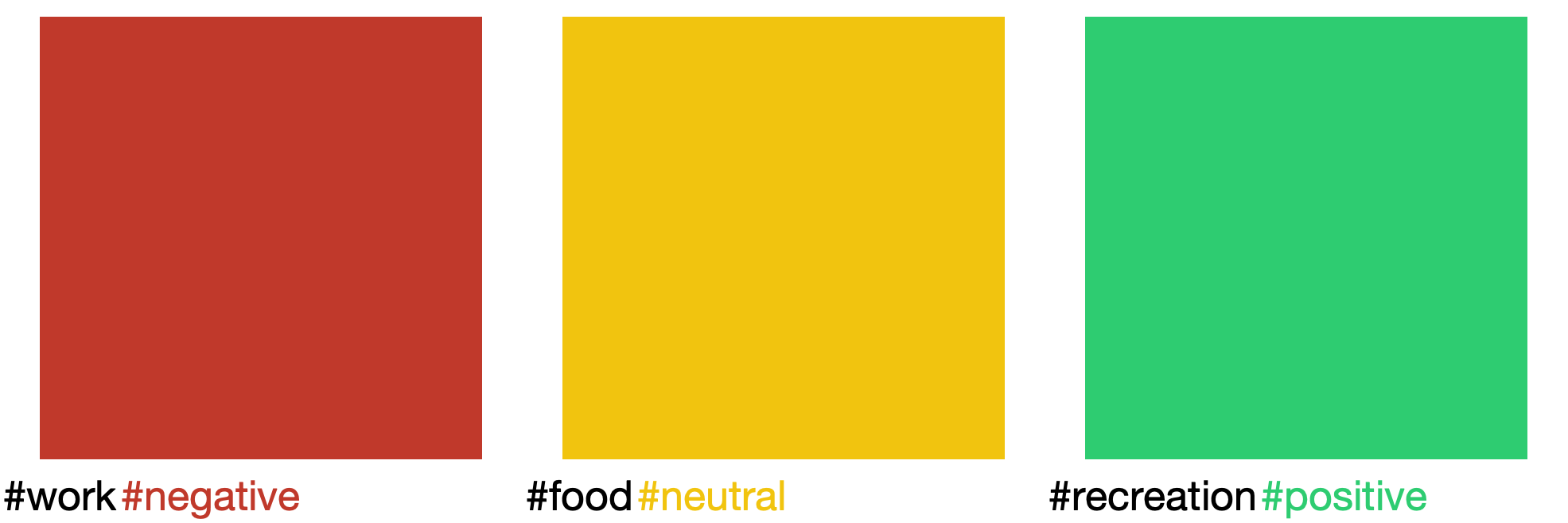}
		\caption*{B6}
	\end{subfigure}
	\begin{subfigure}{0.2\columnwidth}
		\centering
        \includegraphics[width=.96\columnwidth]{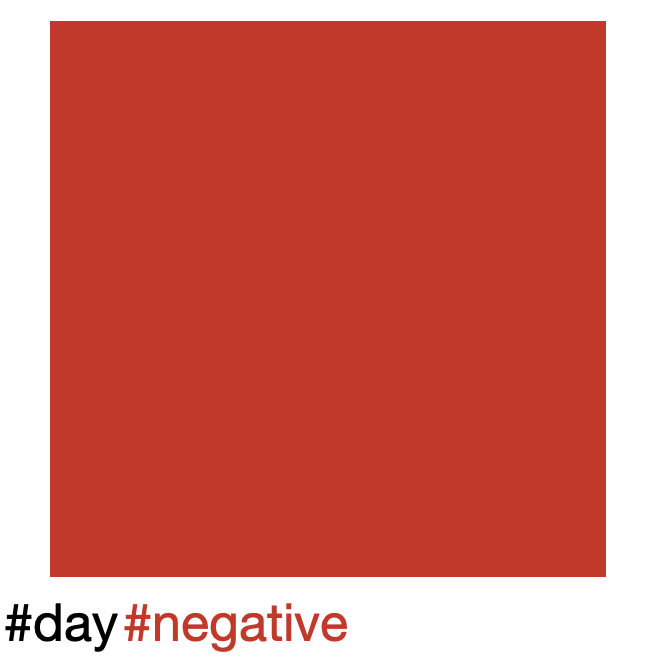}
		\caption*{B7}
	\end{subfigure}
	\caption{Our six proposed visualizations (A1 $\ldots$ A6) and seven comparison baselines (B1 $\ldots$ B7).}
    \label{fig:all_approaches}
    \vspace{-15pt}
\end{figure*}

We train a convolutional neural network designed as an autoencoder,  taking a low-resolution image as its input and predicting a high-resolution version of the same image as its output (generated output shown in A6 in Fig. \ref{fig:all_approaches}). We design our model to perform this form of super-resolution because we want to provide the model a set of colors representing emotions and allow the model to generate creative and stochastic detail --- in other words, we want the model to begin with limited information (colors) and learn higher-resolution artistic representations of the provided colors. Our model consists of three residual blocks ~\cite{he2016deep} encoding a 16x16 input image to feature space and a standard convolutional decoder architecture containing 2D Convolution + BatchNormalization + LeakyReLU blocks producing a 256x256 output image. For our research study, this model is trained on a dataset of 14,621 abstract paintings from WikiArt (downloaded from \code{https://github.com/cs-chan/ArtGAN}), randomly cropped to 256x256 high-resolution ground truths and resized to 16x16 low-resolution inputs. In training, we minimize mean squared error loss between the generated output and the original cropped image. In inference, we randomly populate a 16x16 image with pixel colors corresponding to the provided emotions, producing an output image in the style of abstract art in our target colors. This model can also be trained on different datasets containing artworks from varying artists or artistic movements to produce motifs of diverse styles.

\subsubsection{Carpet} Carpet (A3 in Fig.  \ref{fig:all_approaches}) divides the image into a grid, repeatedly placing parallel lines in each cell of the grid at one of four possible angles and filling in the resulting connected regions with a random color from the set of colors associated with the detected emotions. Users can adjust the thickness and angles of the lines placed and the grid size that the canvas is divided into.

\subsubsection{Circle packing}

In circle packing (A1 in Fig.  \ref{fig:all_approaches}), we fill a blank region of a given shape with circles of differing sizes, each filled with a random color out of the colors associated with the detected emotions. We start with a set of circle radii and the desired number of circles to be placed in the region for each radii, which can be adjusted by users to taste. Starting from the largest size, we sample a random location in the shape region. If a circle can be placed there without any part of the circle falling outside the region or overlapping an existing circle, we draw the circle. If not, we sample another random location and try again until a max number of trials is reached or the circle is successfully placed. This is repeated for the specified number of circles to be placed for each size.

\subsubsection{Glass} Glass (A5 in Fig.  \ref{fig:all_approaches}) attempts to mimic the appearance of stained glass by placing an assortment of icons in the topic shape at differing colors and opacities. By overlapping the canvas region with translucent icons across multiple passes, a random pattern of colors and shapes emerges. Users can customize the number of passes, how densely or sparsely icons are placed, and the distribution of icon sizes.

\subsubsection{Tile} Tile (A4 in Fig.  \ref{fig:all_approaches}) divides the image into a grid, randomly placing a line in each cell along one of two diagonals and filling in the resulting connected regions with randomly chosen colors corresponding to the detected emotions. Users can adjust the grid size, line width, and probability that each one of the two diagonals is picked.

\subsubsection{String Doll}

String Doll (A2 in Fig. \ref{fig:all_approaches}) draws quadratic bezier curves that connect two random points on a blank topical shape's boundary, without the stroke going outside the boundary of the shape. As the control point of the quadratic bezier curve, we take the mid point of the two end points and add zero-mean gaussian noise to it. The standard deviation of the gaussian is set to 20\% of the size of the canvas. The strokes are colored uniformly randomly by one of the colors corresponding to the emotions detected in the user's journal entry for that topic. The width of the stroke is sampled from a distribution of sizes. To add some texture to the visualization, each stroke is overlaid by a stroke that is lighter or darker in color and a quarter of the original stroke's width. Users can adjust the number and width of strokes and the standard deviation of the gaussian controlling the placement of each quadratic bezier curve's control point.

\section{Dataset}
\label{sec:dataset}

We collected a dataset of 500 journal entries from 500 anonymous subjects on Amazon Mechanical Turk (AMT) describing their day, in response to the prompt ``What were salient aspects of your day yesterday? How did you feel about them?'' Figure \ref{fig:teaser} contains an example of one entry from a respondent. Each journal entry contains up to three text samples describing different aspects of the subject's day, referred to as sub-entries henceforth in this paper. We asked subjects to annotate each sub-entry by selecting its associated topics and emotions from a drop down list populated with our set of topics and emotions, serving as ground truth labels for our natural language model evaluation.  Our dataset is available at  \texttt{https://github.com/xaliceli/lemotif}.

\begin{figure}
	\centering
	\begin{subfigure}{1\columnwidth}
		\centering
		\includegraphics[width=1\columnwidth]{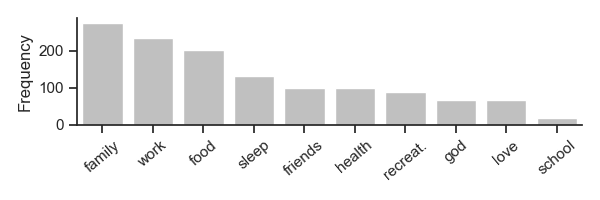}
	\end{subfigure}\vspace{-5pt}
	\begin{subfigure}{1\columnwidth}
		\centering
        \includegraphics[width=1\columnwidth]{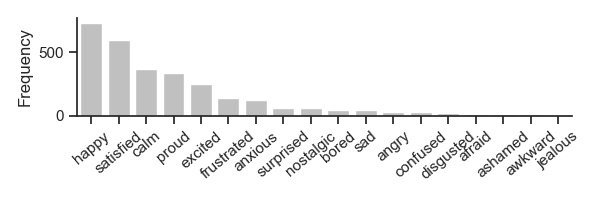}
	\end{subfigure}
	\caption{Distribution of topics and feelings in our dataset.} 
    \label{fig:topics_feelings_dist}
    \vspace{-20pt}
\end{figure}

For entries in our dataset where subjects wrote meaningful responses relevant to the prompt, the mean entry (containing up to three sub-entries) was 507.6 characters (100.6 words) long; on average, each entry included 5.9 emotions and 3 topics. Fig.~\ref{fig:topics_feelings_dist} shows the distribution of topics and feelings subjects chose to talk about. Given that not all 500 respondents wrote three sub-entries and some responses were omitted due to irrelevance to the prompt, 1,473 sub-entries were ultimately used for training and analysis.

Subjects were from the US (to ensure fluent English), had $\ge$95\% approval rating on AMT, and had completed at least $\ge$5000 tasks on AMT in the past. The same qualifications were used for all AMT evaluations discussed in this paper.

\section{Experiments and results}
\label{sec:results}

\subsubsection{Evaluating icon and color choices} 
We showed subjects on AMT our list of 11 topics and a randomly ordered list of the 11 icons shown in Fig.~\ref{fig:icons}. Subjects were asked to assign each icon to exactly one topic. 170 subjects performed this task. Given a topic, the right icon was picked 69\% of the times (mean across subjects), compared to the random chance probability of $\sim$9\%. If we assign a topic to the icon that was picked most often for that topic (majority vote across subjects), the accuracy is 82\%. For a given topic, we sort all icons by how often they were selected across subjects. We find that the right icon falls at rank 1.27 out of 11 (on average across topics). The right icon falls in the top 20\% of the sorted list 91\% of the time across topics, and in the top third of the list 100\% of the time. Overall, subjects appear to find our topic-icon mapping intuitive and natural. 

We ran a similar study to evaluate our feeling-color mapping shown in Fig.~\ref{fig:colors}. This is a more challenging task because (1) icons have descriptive shapes that can be recognized as objects with semantic meaning, while colors are significantly more ambiguous, and (2) there are 18 feelings and colors as opposed to fewer topics and icons. Note that the choice of colors (and icon) being intuitive and natural to users is a bonus, but not a requirement; as seen in Fig.~\ref{fig:teaser}, the topics and feelings are explicitly listed on the motif. 99 subjects participated in this study; because this task was more involved, fewer AMT users elected to participate compared to the topic-icon evaluation. We find that given a feeling, the right color was picked 15\% of the time (mean across subjects). Chance performance would be $\sim$6\%. If we assign a feeling to the color that was picked most often for that feeling (majority vote across subjects), the accuracy is 33\%. For a given feeling, we sort all colors by how often they were selected across subjects. We find that the right color falls at rank 5.28 out of 18 (on average across feelings). The right color falls in the top 20\% of the sorted list 61\% of the time across feelings, and in the top third of the list 67\% of the time. Overall, this shows that despite the mapping being  ambiguous and subjective, subjects do find an intuitive and natural signal in our feelings-color mappings as well. 

\begin{figure}
    \centering
    \includegraphics[width=\columnwidth]{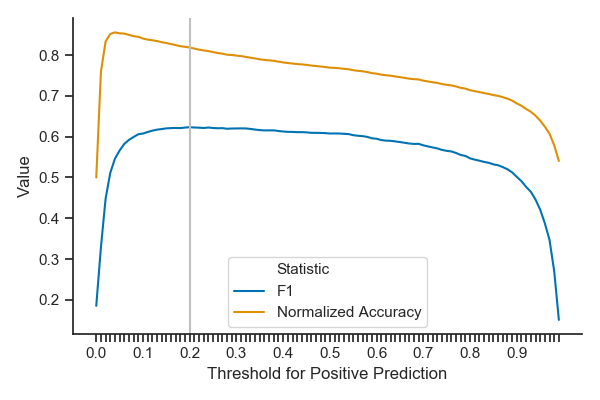}
    \caption{Cross-validation F1 and normalized accuracy statistics by varying probability thresholds used to indicate a positive prediction. Line at 0.2 represents the threshold we select for inference.}
    \label{fig:nlp}
    \vspace{-20pt}
\end{figure}

\subsubsection{Evaluating natural language model}
We trained our NLP model on our text dataset with user-supplied topic-emotion labels serving as ground-truth labels. We performed cross-validation across five train-test splits (80\% train, 20\% test) to calculate normalized accuracy and F1 metrics comparing ground truth versus predicted labels across the full dataset containing 1,473 text samples (sub-entries). Normalized accuracy is the mean between true positive and true negative rates. F1 (also known as F-score) is the harmonic mean of precision and recall. Recall that our NLP model outputs multi-label probability values between 0 and 1. Figure ~\ref{fig:nlp} shows normalized accuracy and F1 scores for various probability thresholds above which a label is counted as a positive classification. At our chosen threshold of 0.2, our model has a normalized accuracy of 82\% and an F1 score of 0.62, compared to random chance values of 50\% and 0.5. Since different thresholds yield similar accuracies, we use 0.2 during inference partially based on experimentation using new and arbitrary input samples. Given that our training set is fairly small, we find that lower thresholds produce undesirably many false positives when exposed to new text samples not from AMT. Additionally, because the distribution of supplied labels is uneven across topics and emotions (as shown in Fig. ~\ref{fig:topics_feelings_dist}), we acknowledge that our model may not perform well on new samples referring to topics or emotions underrepresented in our dataset.

\subsubsection{Evaluating creative motifs}

The generated motifs should (1) separate the topical sources of emotions, (2) depict these sources visually, (3) depict the emotions visually, and (4) be creative and attractive. To evaluate this hypothesis, we design several baselines that allow us to measure the role of each factor. We strip away one factor at a time to derive our various baselines. To keep the number of baselines manageable, we create these baseline versions only for circle packing (Fig.~\ref{fig:all_approaches} A1) and string doll (Fig.~\ref{fig:all_approaches} A2). 
\begin{itemize}
\item
We start with our generated motif and remove the shape depiction, retaining the creative design, color depictions, and separate topic depictions. We replace each icon shape with a square whose contents are rendered according to our visualization styles.
This gives us two baselines (B1 and B2) in Fig.~\ref{fig:all_approaches}.
\item 
We can also start with our generated motifs and remove the creative design, while maintaining the shape and color depictions, as well as the topic breakdown. We color each shape
with solid colors associated with the feelings mentioned for that topic.  
This gives us B3 in Fig.~\ref{fig:all_approaches}.
\item 
We can now remove the shape information from the above baseline, and depict squares (instead of icons) for each topic colored in with solid colors (no creative aspect). 
This gives us B4 in Fig.~\ref{fig:all_approaches}.
\item
We can start with B3 and remove the detailed color information. 
Instead of using a color for each of the 18 feelings, we use just three colors: red, yellow, and green to depict negative, neutral or positive feelings. We mapped afraid, angry, anxious, ashamed, disgusted, frustrated, jealous and sad to negative; awkward, bored, calm, confused, nostalgic and surprised to neutral; and excited, happy, proud and satisfied to positive. We use the majority label across reported feelings to pick a color for that topic. 
This gives us B5 in Fig.~\ref{fig:all_approaches}.
\item
We can remove shape information from the above baseline to have squares colored in either red, yellow or green  representing each topic. 
This gives us B6 in Fig.~\ref{fig:all_approaches}.
\item 
Finally, we can remove the topic breakdown from the above baseline and have the entire day depicted as a red, yellow, or green square based on the most common label across reported feelings for the day. 
This gives us B7 in Fig.~\ref{fig:all_approaches}. As mentioned in the related work section, this mimics an existing app (Life Calendar) that shows a single colored dot for every week in the year. 
\end{itemize}

To start, we combine these seven baselines with two of our proposed visualizations (Fig.~\ref{fig:all_approaches} A1 and A2), giving us nine approaches to compare for this evaluation; later on we will compare all six visualization styles described in the Approaches section against each other and a smaller set of three baselines. We generate these visualizations for a subset of 100 journal entries from our dataset, using user-supplied ground-truth topic-emotion labels; each entry contains three sub-entries and their corresponding motifs. For parameters users can vary, such as line width and spacing, we set their values based on what we found most visually appealing and representative of each style's overall design. We conduct pairwise evaluations on AMT. We show subjects a journal entry from our dataset, and all $\binom{9}{2} = 36$ pairs of visualizations. For each pair, we ask subjects ``If you were using a journaling tool or app that automatically produced a visual summary of your day, which one of these visualizations would you prefer?'' 936 unique subjects participated in this study, each providing us a rating for the 36 pairs for a single journal entry. Each journal entry was evaluated by 6 to 17 subjects, with an average of 9.4 and mode of 10.

By comparing pairs of the proposed approaches, we can evaluate the role of the four visual factors listed above. How often subjects pick B6 over B7 reveals how important it is for the motif to have a breakdown across topics. Similarly, comparing B5 to B6, B3 to B4, A1 to B1, and A2 to B2, indicates the importance of a topic being depicted by a shape as opposed to a generic square. Comparing B3 to B5, and B4 to B6, indicates the importance of each feeling being depicted by a nuanced rather than coarse color for negative, neutral, and positive feelings. Comparing A1 to A2 indicates which of the two creative motifs subjects prefer. We find that subjects prefer circle packing (A1) to string doll (A2)  72\% of the time. We focus our evaluation of the creative aspect on A1. Comparing A1 to B3 and B1 to B4 reveals how much subjects prefer creative designs.

\begin{figure}[t!]
    \centering
    \includegraphics[width=\columnwidth]{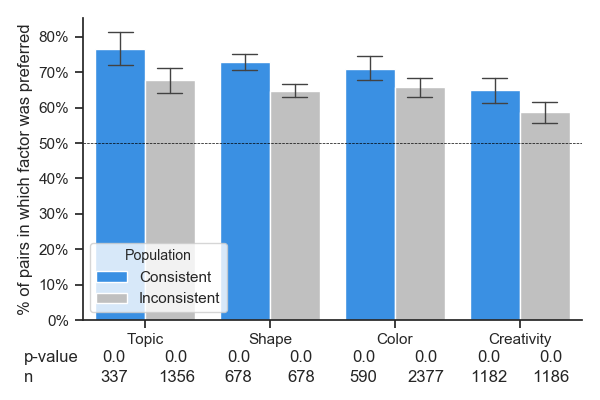}
    \caption{Percentage of times subjects prefer a visualization with the four factors over corresponding baselines, for subjects who were consistent across their pairwise preferences and those who were not. P-value is from a one-sample t-test compared to null hypothesis of 50\%, i.e. random chance (shown as dashed line). N reflects the number of pairs in which each relevant comparison was performed. Error bars represent 95\% confidence intervals.}
    \label{fig:consistency}
    \vspace{-20pt}
\end{figure}

In Fig.~\ref{fig:consistency}, for each of the four factors, we show how often a visualization with that factor is preferred over a corresponding visualization without that factor (as described above). We show these statistics separately for subjects who were consistent in their preferences vs. those that had some contradictory preferences. Recall that we had each subject report their preferences for all $\binom{9}{2} = 36$ visualization pairs for a single journal entry. We can check whether the pairwise preferences reported are consistent across the board or not (if a $>$ b and b $>$ c, then a should be $>$ c). Presumably, subjects who provide consistent preferences are likely to be doing the task more carefully and/or have more clear preferences. We find that 36\% of our subjects were perfectly consistent across the 36 pairwise comparisons. Across the board in Fig.~\ref{fig:consistency}, the four factors are preferred, especially for subjects who were consistent in their responses. 

\begin{figure*}[ht]
	\centering
	\begin{subfigure}{1\columnwidth}
		\centering
		\includegraphics[width=1\columnwidth]{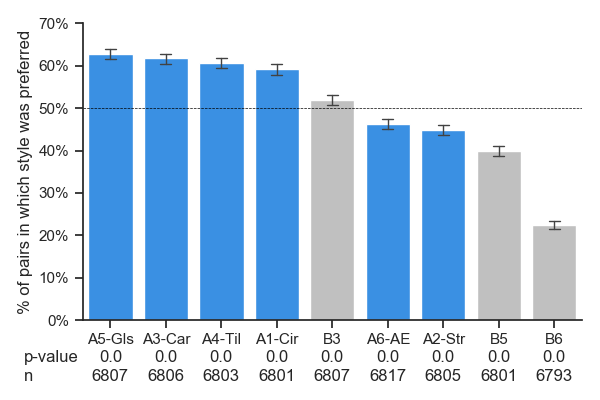}
		\vspace{-10pt}
		\caption{Percent of pairs in which style was preferred. P-value is from a one-sample t-test compared to null hypothesis of 50\%. N reflects the number of pairs in which each style was compared.}
	\end{subfigure}
    \hspace{20pt}
	\begin{subfigure}{1\columnwidth}
		\centering
        \includegraphics[width=1\columnwidth]{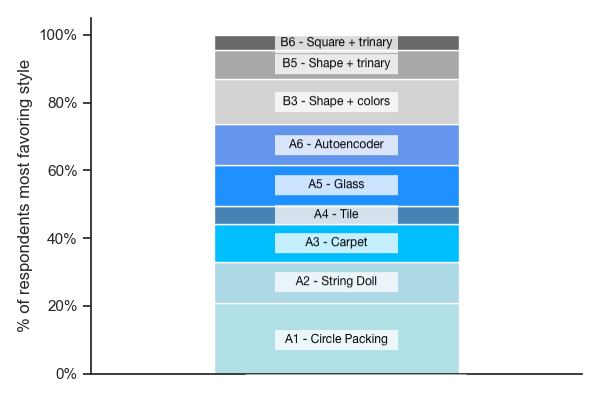}
        \vspace{-10pt}
        \caption{Percent of respondents who favored each style above all other styles. "Favorite style" is the style which was chosen most times across all pairs seen for each respondent.}
	\end{subfigure}
	\caption{Preferences across visualizations. Creative styles are shown in blue and baseline comparisons are shown in gray.} 
	\label{fig:all_vis_pref}
    \vspace{-20pt}
\end{figure*}

\subsubsection{Evaluating additional visualization styles}

Having established that our four creative factors are generally preferred by human subjects, we next evaluate all six visualization styles (A1-A6 in Fig.~\ref{fig:all_approaches}) against a smaller set of three baselines: B3 (topical shapes and emotional colors with no additional creative style), B5 (topical shapes and positive-neutral-negative colors), and B6 (squares and positive-neutral-negative colors). Similar to the first evaluation performed, we generate 36 visualization pairs and ask respondents to select the style they prefer. 854 unique subjects participated in this study.

Fig.~\ref{fig:all_vis_pref} shows user preferences across all six visualization styles and three baseline comparisons. Overall, the most preferred styles were the creative visualizations, consistent with what we saw in the prior evaluation. The one baseline comparison that performed comparably to random chance was B3, which includes both topical shapes and our full set of 18 emotional colors; though this baseline was not intentionally designed as a creative style, one could argue that placing colors in equally distributed regions \textit{is} a creative visualization. After all, there are entire artistic movements such as color field painting with similarly "flat" aesthetics. We also note that, when evaluating which style was each respondent's favorite (defined as the style that was most frequently preferred for each respondent), preferences are widely distributed across styles. For example, even though the autoencoder was preferred fewer than 50\% of the time overall, 12\% of respondents preferred it above all other styles, comparable to the 12\% of respondents who most favored the glass visualization which scored highest in pairwise comparisons. The diversity of preferences highlights the personal nature of aesthetics and how the act of choosing a motif to use can be a creative decision in and of itself.

\subsubsection{Evaluating engagement}

The real evaluation of a system like Lemotif is how users would engage with it --- would users journal more regularly, feel more creative, and/or gain actionable insights from their motifs? Such a longitudinal evaluation is outside the scope of this paper. As a proxy, we ran two surveys on AMT. The first survey (study S1 with 100 unique and valid responses) described the concept of Lemotif to subjects and showed example circle packing motifs for reference. The second survey (study S2 with 99 unique and valid responses) directed subjects to a web demo asking them to write three pieces of text about their day and generated motifs in all six visualization styles based on labels automatically detected from their entries. Note that S2 evaluates our entire system end-to-end on free-form entries.

Table ~\ref{tab:eval_survey} shows the percentage of respondents agreeing with each statement. In both studies, a majority of subjects stated they would use an app like Lemotif, that such an app would make journalling more enjoyable, that they would write more regularly with an app like Lemotif, and that the motifs "get their creative juices flowing." 71\% of respondents who used the end-to-end demo agreed the motifs were representative of what they wrote, further affirming that our NLP model effectively extracts topic-emotion labels and that our mapping of abstract concepts to visual representations feels intuitive to human subjects. Between S1 and S2, responses to the metrics shown are comparable, with the exception of "more enjoyable" and "get creative juices flowing" receiving lower scores in the end-to-end demo. Since S1 only describes Lemotif to users, they are free to imagine an ideal user interface. Moreover, assessment of the end-to-end demo would also suffer from errors in the NLP model, which is not perfect. We posit that a full app with attention to user experience design and our full set of customization options would likely score higher than S2 currently.

\begin{table}[ht]
\caption{Survey responses to Lemotif app}
\begin{center}
\begin{tabular}{ |l|c|c| } 
 \hline
  \textbf{Question} & \textbf{\%Yes (S1)} & \% \textbf{Yes (S2)} \\ 
 \hline
 Representative of entry? & NA & 71\% \\ 
 \hline
 Would use? & 59\% & 56\% \\ 
 \hline
 Make more enjoyable? & 68\% & 59\% \\ 
 \hline
 Would write more regularly? & 61\% & 61\% \\ 
 \hline
 Get creative juices flowing? & 59\% & 51\% \\ 
 \hline
\end{tabular}
\end{center}
\label{tab:eval_survey}
\end{table}

In the full demo, respondents were also asked to select their favorite visualization style out of all six presented in randomized order. Table \ref{tab:demo_fav} shows the percentage of respondents selecting each style as their favorite out of 84 respondents who answered this question. Similar to our other evaluations, we see that no one style is dominantly favored.

\begin{table}[ht]
\caption{Percentage of respondents choosing style as favorite}
\begin{center}
\begin{tabular}{ |l|c| } 
 \hline
  \textbf{Style} & \textbf{\% Favorite} \\ 
  \hline
  Autoencoder (A6) & 14\% \\ 
   \hline
 Carpet (A3) & 19\% \\ 
  \hline
 Circle Packing (A1) & 23\% \\ 
  \hline
 Glass (A5) & 8\% \\ 
   \hline
 Tile (A4) & 25\% \\ 
   \hline
 String Doll (A2) & 11\% \\ 
 \hline
\end{tabular}
\end{center}
\label{tab:demo_fav}
\end{table}

Overall across our multiple evaluations, we see that (1) a majority of subjects find our visual representation of abstract concepts intuitive, (2) our NLP model extracts accurate labels for a majority of entries, (3) a majority of subjects prefer our motif designs over corresponding baselines, and (4) a majority of prospective users consider Lemotif a useful system that would increase their likelihood to journal and enjoyment of journaling.

\section{Future Work}

Future work involves developing Lemotif into an app that allows users to accumulate their entries over time and view temporal (weekly, monthly, etc.) summaries. The app could allow for custom mappings, such as allowing users to specify the name of their job so the NLP model always identifies it as "work," or correct the detected topics and emotions so that over time the model learns the user's personal life and writing style. Training our model with more data containing more diverse labels would also likely improve its accuracy.

Additional visualization styles are possible given the diversity of generative art. Our autoencoder model would likely improve with architectural changes, adversarial discriminator loss (like a GAN), and hyperparameter tuning. With a sufficiently large and annotated dataset, a conditional GAN could be trained that takes in color labels directly rather than as low-resolution images. Multiple models could be trained on different artists and artistic movements. Within an app system, users could provide feedback on generated motifs they like more or less, further training the image models to the user's own taste. Additional input dimensions like the intensity of emotion could be incorporated, such that stronger emotions appear more saturated.

\section{Conclusion}
In summary, we present Lemotif. It takes as input a text-based journal entry indicating what aspects of the user's day were salient and how they made them feel and generates as output a motif -- a creative abstract visual depiction -- of the user's day. As a visual journal used over periods of time, Lemotif aims to make associations between feelings and parts of a user's life more apparent, presenting opportunities to take actions towards improved emotional well being. 

Lemotif is built on five underlying principles: (1) separate out the sources of emotions, (2) depict these sources visually, (3) depict these emotions visually, (4) generate visualizations that are creative and attractive, and (5) identify and visualize detected topics and emotions automatically using machine learning and computational methods. We verify via human studies that each of the first four factors contributes to the proposed motifs being favored over corresponding baselines; accuracy and F1 metrics indicate the NLP model greatly outperforms random chance. We also find that subjects are interested in using an app like Lemotif and consider the generated motifs representative of their journal entries.

\subsubsection{Acknowledgments}                      
\small{Thanks to Ayush Shrivastava, Gauri Shri Anantha, Abhishek Das, Amip Shah, Sanyam Agarwal, Eakta Jain, and Geeta Shroff for participating in an earlier version of this study. Special thanks to Abhishek Das for useful discussions and feedback. At Facebook AI Research, we understand that researching a topic like emotion is nuanced and complicated. This work does not research what causes emotional well being (or not). It does not mine Facebook data to extract emotions, or use Facebook data or the Facebook platform in any other way. It simply generates visualizations based on topics and emotions reported by subjects explicitly electing to participate in our study, and analyzes which visualizations subjects prefer. Creative applications of AI are a powerful avenue by which AI can collaborate with humans for positive experiences. This work is one (small) step in that direction.}

\bibliographystyle{iccc}
\small
\bibliography{iccc}

\end{document}